\documentclass{article}
\usepackage{amsmath,amssymb,epsfig}

\begin{document}

\thispagestyle{empty} \vspace*{1cm}

\begin{center}
{\LARGE Point-like topological defects in bilayer Quantum Hall systems}

{\LARGE \ }

\vspace{8mm}

{\large Gerardo Cristofano\footnote{{\large {\footnotesize Dipartimento di
Scienze Fisiche,}\textit{\ {\footnotesize Universit\'{a} di Napoli
``Federico II''\ \newline
and INFN, Sezione di Napoli}-}{\small Via Cintia - Compl.\ universitario M.
Sant'Angelo - 80126 Napoli, Italy}}}, Vincenzo Marotta\footnotemark[1]  , }

{\large Adele Naddeo\footnote{{\large {\footnotesize Dipartimento di Scienze
Fisiche,}\textit{\ {\footnotesize Universit\'{a} di Napoli ``Federico II''
\newline
and Coherentia-INFM, Unit\`{a} di Napoli}-}{\small Via Cintia -
Compl. universitario M. Sant'Angelo - 80126 Napoli, Italy}}} },
{\large Giuliano
Niccoli\footnote{{\large {\footnotesize Ecole Normale Superieure de Lyon-}%
{\small 46 Allee d'Italie - 69364 Lyon cedex 07, France}}} }

{\small \ }

\textbf{Abstract\\[0pt]
}
\end{center}

\begin{quotation}
Following a suggestion given in \cite{noi}, we show how a bilayer Quantum
Hall system at fillings $\nu =\frac{m}{pm+2}$ can exhibit a point-like
topological defect in its edge state structure. Indeed our CFT theory for
such a system, the Twisted Model (TM), gives rise in a natural way to such a
feature in the twisted sector. Our results are in agreement with recent
experimental findings \cite{deviatov1} which evidence the presence of a
topological defect in the bilayer system.

\vspace*{0.5cm}



{\footnotesize Work supported in part by the European Communities Human
Potential}

{\footnotesize Program under contract HPRN-CT-2000-00131 Quantum
Spacetime\newpage } \setcounter{page}{2}
\end{quotation}

The quantum Hall effect (QHE) is one of the most remarkable
many-body phenomena discovered in the last twenty-five years
\cite{prangegirvin}\cite{tp}\cite {perspectives}. It takes place
in a two-dimensional electron gas formed in a quantum well in a
semiconductor host material and in the presence of a very high
magnetic field \cite{dorda} \cite{stormer}, as a result of the
commensuration between the number of electrons $N$ and the number
of flux quanta $N_{\Phi }$. The electrons condense into distinct
and highly non-trivial ground states (`vacua') formed at each
integer (IQHE) \cite {laughlin1} or rational fractional value
(FQHE)\cite{laughlin2} of the filling factor $\nu
=\frac{N}{N_{\Phi }}$. In particular at fractional fillings
quasiparticles with fractional charge and statistics emerge, and
new kinds of order parameters are considered \cite{ODLRO}. The
essential feature of such peculiar states is the existence of an
excitation gap, so that the electron fluid is incompressible and
flows rigidly past impurities in the sample without
dissipation. As a result the conductivity tensor takes the universal form $%
\sigma^{xx}= \sigma^{yy}=0$, $\sigma^{xy}=-\sigma^{yx}=
\nu\frac{e^{2}}{h}$. The presence of disorder in the samples is
crucial for the observed phenomenology in order to localize
topological defects and prevent dissipation.

Since the first observations of the FQHE \cite{stormer}
considerable experimental progress has been made in performing
measurements with samples of higher mobility under stronger
magnetic fields and at lower temperatures.

The experimental results relative to the sequence of filling factors $\nu=%
\frac{p}{(2p\pm 1)}$ and $\nu=\frac{p}{(4p\pm 1)}$ have given strong support to
the idea first suggested by Jain \cite{jain} of looking at the FQHE for
electrons as a manifestation of the integer effect (IQHE) for composite
fermions, obtained by attaching to each electron an even number of flux
units opposite to the external magnetic field. In fact the most prominent
Hall plateaux have been observed at the fillings of the principal sequence $%
\frac{p}{(2p\pm 1)}$ and the energy gaps measured for this sequence have
been found to correspond to the cyclotron energies relative to the reduced
magnetic field $B-B_{1/2}$ \cite{du}. Furthermore Halperin, Lee and Read
\cite{lee}, following a point of view closely related to Jain's approach,
stressed the role of the state at $\nu =\frac{1}{2}$, computing an anomaly in
surface acoustic wave propagation in agreement with experimental results \cite{willett}.

The experimental evidence of a Hall plateau at filling
$\nu=\frac{5}{2}$ focused the physicists attention to plateaux
which do not fall into the hierarchical scheme \cite{jain}. To
such an extent a pairing picture, in which pairs of spinless or
spin-polarized fermions condense, has been proposed \cite{MR} for
the non-standard fillings $\nu=\frac{1}{q}$, $q>0$ and even. As a
result the ground state has been described in terms of the
Pfaffian (the so called Pfaffian state) and the non-Abelian
statistics of the fractional charged excitations evidenced
\cite{MR}\cite{FNTW}\cite{MIR}\cite{FNS}.

Recent technological progress in molecular beam epitaxy techniques
has led to the ability to produce pairs of closely spaced
two-dimensional electron gases. Since then such bilayer quantum
Hall systems have been widely investigated theoretically as well
as experimentally \cite{teoria, esperim}. Strong correlations
between the electrons in different layers lead to new physical
phenomena involving spontaneous interlayer phase coherence. In
particular a spontaneously broken $U(1)$ symmetry \cite{Zee} has
been discovered and identified and many interesting properties of
such systems have been studied: the Kosterlitz-Thouless
transition, the zero resistivity in the counter-current flow, a
DC/AC Josephson-like effect in interlayer tunneling as well as the
presence of a gapless superfluid mode \cite{Zee1}\cite {girvin1}.
Indeed, when tunneling between the layers is weak, the quantum
Hall bilayer state can be viewed as arising from the condensation
of an excitonic superfluid in which an electron in one layer is
paired with a hole in the other layer. The uncertainty principle
makes it impossible to tell which layer either component of this
composite boson is in. Equivalently the system may be regarded as
a ferromagnet in which each electron exists in a coherent
superposition of the ''pseudospin'' eigenstates, which encode the
layer degrees of freedom \cite{YangMoon}\cite{girvin1}. The phase
variable of such a superposition fixes the orientation of the
pseudospin magnetic moment and its spatial variations govern the
low energy excitations in the system.

Since Halperin work \cite{Halperin} the concept of edge states was
introduced in order to describe transport phenomena in two dimensional
electron systems. They arise in a quantized magnetic field at the
intersections of the Fermi level with different Landau levels, which are
bent up by the edge potential. In particular the formation of a topological
defect has been predicted to occur when two edge states with different spins
locally switch their positions and thus cross each other at two or more
points \cite{edge1}. More interesting features take place in the transport
properties of bilayer systems when also pseudospin (related to the layer
index) is involved \cite{girvin1}\cite{sarma}. Recently the presence of edge
state crossings and thus of topological defects has been experimentally
evidenced in such systems in a quasi-Corbino geometry \cite{deviatov2} at
filling $\nu =3$ \cite{deviatov1} by means of a selective population
technique. In particular the application of a suitable gate voltage $V_{g}$
and of a magnetic field drives the bilayer in different pseudospin states in
the gated and ungated regions, so producing a crossing of the edge states
which has been detected in the transport characteristics. More precisely the
gated region has filling $g$ while the ungated region is at filling $\nu -g$.

The net result is a linear $I-V$ characteristics for the electric transport
between two different edges. Because the gate-gap width is smaller than the
characteristic equilibration lengths in such a transport between the edge
states, it has been argued that a defect must be present, which couples
different edge states but only with the same spin in the gate-gap. Such a
picture can be destroyed by an in-plane magnetic field component which
washes out the above crossing; the $I-V$ curves become then strongly
non-linear as a signal of the merging of a tunneling process. All the above
features in the $I-V$ characteristics appear to be the fingerprints of the
presence of a topological defect induced by the different pseudospin
configurations in bilayer quantum Hall systems \cite{deviatov1}.

In this letter we address theoretically the issue of the presence of
topological defects in the Conformal Field Theory (CFT) description of the
edge states of bilayer quantum Hall systems in a wide class of filling
factors, and in particular the paired states ones, in the framework of our
TM approach \cite{noi}. In particular we show how such a feature arises in a
very natural way in the twisted sector of our theory, as a result of the $m$%
-reduction technique \cite{cgm1, cgm4}. The transport properties of bilayer
systems will be investigated by studying the properties under magnetic
translations of the characters of the different sectors, which describe its
different non perturbative ground states. We point out that the results of
this letter are very general and are relevant for different areas of
condensed matter systems at low dimensions. In particular it has been shown
that there is a close relation between the existence of topological defects
and flux fractionalization in fully frustrated Josephson junction ladders
\cite{JJL}. Furthermore topological defects have been also introduced in the
description of dissipation in systems with magnetic impurities \cite{noi}.

The $m$-reduction technique is based on the simple observation that for any
CFT (mother) exists a class of sub-theories parameterized by an integer $m$
with the same symmetry but different representations. The resulting theory
(daughter) has the same algebraic structure but a different central charge $%
c_{m}=mc$. To obtain the generators of the algebra in the new
theory we need to extract the modes which are multiple of the
integer $m$. These can be used to reconstruct the primary fields
of the daughter CFT. This technique can be generalized and applied
to any extended chiral algebra which includes the Virasoro one.
Indeed the $m$-reduction preserves the commutation relations
between the algebra generators but modifies the central extension
(i.e. the level for the WZW models). In particular this implies
that the number of primary fields gets modified. Its application
to the QHE arises by the incompressibility of the Hall fluid
droplet at the plateau, which implies its invariance under the
$W_{1+\infty }$ algebra at the different fillings, and by the
property of the $m$-reduction procedure to obtain a daughter CFT
with the same\ $W_{1+\infty }$ invariance property of the mother
theory. Thus the $m$-reduction furnishes automatically a mapping
between different incompressible plateaux of the quantum Hall
fluid (QHF).

The general characteristics of the daughter theory is the presence of
twisted boundary conditions (TBC) which are induced on the component fields.
It is illuminating to give a geometric interpretation of that in terms of
the covering on a $m$-sheeted surface or complex curve with branch-cuts, see
Fig. 1.
\begin{figure}[ht]
\centering\includegraphics*[width=0.3\linewidth]{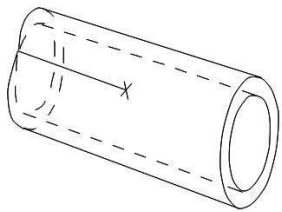}
\caption{The boundaries of the 2-covered cylinder can be viewed as different
configurations of the QHF edges described by the 2-reduced CFT.}
\label{figura1}
\end{figure}
Indeed the fields which are defined on the left boundary have TBC while the
fields defined on the right one have periodic boundary conditions (PBC). We
point out that fields with TBC describe elegantly the crossing between the
layers as a consequence of the presence of a branch-cut. We find different
sectors on the torus corresponding to different boundary conditions on the
cylinder. Finally we recognize the daughter theory as an orbifold of the
usual CFT describing the QHF at a given plateau. The two sheets simulate the
two-layers system and the branch cut represents TBC which emerge from the
interaction with a localized defect on the edge. This is a key feature of
our construction, as we will point out in the following.

In order to see how the $m$-reduction procedure works on the plane \cite
{cgm1} and on the torus \cite{cgm4} and how it gives rise to the edge state
coupling via a topological defect, let us focus on the paired states
fillings in the special $m=2$ case since we are interested in a system
consisting of two parallel layers of $2D$ electrons gas in a strong
perpendicular magnetic field. The filling factor $\nu ^{(a)}=\frac{1}{2p+2}$
is the same for the two $a=1$, $2$ layers while the total filling is $\nu
=\nu ^{(1)}+\nu ^{(2)}=\frac{1}{p+1}$. We point out that our results can be
generalized to any bilayer system. The simplest abelian quantum Hall state
in the disc topology is written as a generalization of the analytic part of
the Laughlin wave function \cite{Halperin}:
\begin{equation}
f\left( z_{i}^{(a)}\right) =\prod_{a=1,2}\prod_{i<j}\left(
z_{i}^{(a)}-z_{j}^{(a)}\right) ^{2+p}\prod_{i,j}\left(
z_{i}^{(1)}-z_{j}^{(2)}\right) ^{p} ;  \label{halp}
\end{equation}
in particular, for $p=0$ it describes the bosonic $220$ state and, for $p=1$%
, the fermionic $331$ one. The CFT description for such a system
can be given in terms of two compactified chiral bosons $Q^{(a)}$
with central charge $c=2$. A similar result can be obtained for
filling $\nu ^{\left( a\right) }=1/(2p+1)$.

In order to construct the fields $Q^{(a)}$ for the TM, let us
start from the bosonic filling $\nu =1/2(p+1) $, described by a
CFT with $c=1$ in terms of a scalar chiral field $Q$
compactified on a circle with radius $R^{2}=1/\nu =2(p+1)$ (or its dual $%
R^{2}=2/(p+1)$). It is explicitly given by:
\begin{equation}
Q(z)=q-i\,p\,lnz+i\sum_{n\neq 0}\frac{a_{n}}{n}z^{-n}  \label{modes}
\end{equation}
with $a_{n}$, $q$ and $p$ satisfying the commutation relations $\left[
a_{n},a_{n^{\prime }}\right] =n\delta _{n,n^{\prime }}$ and $\left[ q,p%
\right] =i$. From such a CFT (mother theory), using the $m$-reduction
procedure, which consists in considering the subalgebra generated only by
the modes in eq. (\ref{modes}) which are a multiple of the integer $m$, we
get a $c=2$ orbifold CFT (daughter theory, i.e. the TM) which describes the
LLL dynamics. Then the fields in the mother CFT can be organized into
components which have well defined transformation properties under the
discrete $Z_{2}$ (twist) group, which is a symmetry of the TM. Its primary
fields content can be expressed in terms of a $Z_{2}$-invariant scalar field
$X(z)$, given by
\begin{equation}
X(z)=\frac{1}{2}\left( Q^{(1)}(z)+Q^{(2)}(-z)\right) ,  \label{X}
\end{equation}
describing the electrically charged sector of the new filling, and a twisted
field
\begin{equation}
\phi (z)=\frac{1}{2}\left( Q^{(1)}(z)-Q^{(2)}(-z)\right) ,  \label{phi}
\end{equation}
which satisfies the twisted boundary conditions $\phi (e^{i\pi }z)=-\phi (z)$
and describes the neutral sector \cite{cgm1}. Such TBC signal the presence
of a topological defect which couples, in general, the $m$ edges in a $m$%
-layers system. In the bilayer system ($m=2$) we get a crossing between the
two edges as sketched in Fig. 2.

\begin{figure}[ht]
\centering\includegraphics*[width=0.8\linewidth]{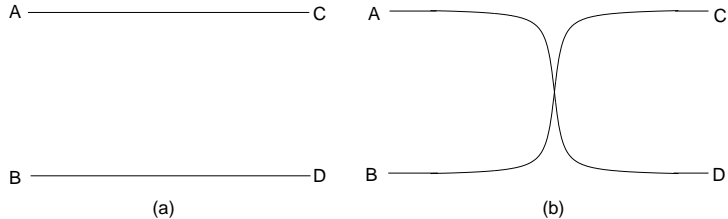}
\caption{The bilayer system, (a) without the topological defect (PBC), (b)
with the topological defect (TBC).}
\label{figura2}
\end{figure}

The chiral fields $Q^{(a)}$, defined on a single layer $a=1$, $2$, due to
the boundary conditions imposed upon them by the orbifold construction, can
be thought of as components of a unique \textquotedblleft boson
\textquotedblright\ defined on a double covering of the disc (layer) ($%
z_{i}^{(1)}=-z_{i}^{(2)}=z_{i}$). As a consequence of such a construction
the two layers system becomes equivalent to one-layer QHF and the $X$ and $%
\phi $ fields defined in eqs. (\ref{X}) and (\ref{phi}) diagonalize the
interlayer interaction. In particular the $X$ field carries the total charge
with velocity $v_{X},$ while $\phi $ carries the charge difference of the
two edges with velocity $v_{\phi }$ i.e. no charge, being the number of
electrons the same for each layer (balanced system).

The TM primary fields are composite operators and, on the torus, they are
described in terms of the conformal blocks (or characters).

The defect is a topological one and in our formalism is induced by
the different isospin configurations on the two layers, which
naturally result from our $m$-reduction procedure. The effect of a
topological defect in a QHF has been recently evidenced in
experimental findings \cite {deviatov1}, as we will show in the
following. In the presence of a localized defect two phenomena can
take place. A tunneling of edge quasi-particles at point $x_{0}$,
described by\ a boundary term Hamiltonian such as:
\begin{equation}
H_{P}=-t_{P}\cos \left( Q^{(1)}-Q^{(2)}\right) \delta \left( x_{0}\right) .
\label{tunnel}
\end{equation}
A second mechanism producing a current flow between the two edges can be
addressed to a localized crossing of the edges,\ which can be represented by
a boundary term:
\begin{equation}
H_{\beta }=\beta \left( Q^{(1)}\partial _{t}Q^{(2)}-Q^{(2)}\partial
_{t}Q^{(1)}\right) \delta \left( x_{0}\right) ,  \label{magterm}
\end{equation}
where $\beta =0(1/2)$ for PBC (TBC) respectively (see Fig. 2). The full
Hamiltonian can be written as:
\begin{eqnarray}
H &=&\frac{1}{2}\sum_{a=1,2}\left[ \left( \Pi ^{(a)}\right) ^{2}+\left(
\partial _{x}Q^{(a)}\right) ^{2}\right] +H_{P}+H_{\beta }  \nonumber \\
&+&eV\partial _{t}\left( Q^{(1)}-Q^{(2)}\right) ,  \label{fullham}
\end{eqnarray}
where $\Pi ^{(a)}$ is the momentum conjugate to $Q^{(a)}$. We recognize a
kinetic term for the two bosonic fields $Q^{(a)},a=1,2$, a boundary
tunneling term which implements the locally applied gate voltage $%
V_{g}=t_{P}\delta \left( x_{0}\right) $, a boundary magnetic term \cite
{maldacena} which couples the two fields introducing a topological defect
(see ref. \cite{noi} for details) and a voltage switching term between the
two layers. The last term contains an irrelevant operator, so it doesn't
change the central charge: it behaves as a boundary condition changing
operator allowing for the flow from a boundary state to another one.
Introducing the charged and neutral fields $X$ and $\phi $ defined in eqs. (%
\ref{X}) and (\ref{phi}) we clearly see that the last term in the
Hamiltonian is proportional to the neutral current, so it contributes to
unbalance the system. Therefore edge-crossing can be described by a TBC on
the $\phi $ field induced by the boundary magnetic term of eq. (\ref{magterm}%
).

The transport properties of such a system can be investigated by the
application of different chemical potentials between the terminals of Fig.
2, that we represent by the matrix $V=\left(
\begin{array}{cc}
V_{AC} & V_{AD} \\
V_{BC} & V_{BD}
\end{array}
\right) $ with entries $V_{IJ}$, the potentials between the $I$ and $J$
terminals. Let us consider the following two cases, the one in which the
transport of electrons is on the two independent edges through the points $%
A-C-A$, $B-D-B$ in the non crossed case (PBC, see Fig. 2a) and the one in
which the transport is through the points $A-D-B-C-A$ in the crossed edge
case (TBC, see Fig. 2b). In both cases there is no tunneling ($t_{p}=0$) and
they correspond respectively to the diagonal (i.e. $V_{AD}=V_{BC}=0$) and to
the anti-diagonal (i.e. $V_{AC}=V_{BD}=0$) configurations. In a closed
geometry, such as that of a torus, they can be induced by adiabatic magnetic
flux insertion through a cycle of the torus (i.e. $A$ or $B$ cycle). For
example, by inserting a flux quantum $\frac{hc}{2e}$ through the cycle $A$,
an electromotive force is induced along it with a consequent transport of an
electron along the $B$ cycle. In the torus topology the transport properties
can be precisely described in terms of the action of magnetic translations
on the conformal blocks of the untwisted and twisted sector respectively.
Their explicit description can be realized by standard calculations on the
characters of the TM given in refs \cite{cgm4}. In this letter we just
recall that the characters are given in terms of opportune Jacobi theta
functions with characteristics $\theta \left[
\begin{array}{c}
\lambda \\
0
\end{array}
\right] \left( qw^{\left( i\right) }|2q\tau \right) $, where $\tau $ is the
modular parameter of the torus and $w^{\left( i\right) }=x^{\left( i\right)
}+y^{\left( i\right) }\tau $ is the torus coordinate of the electron ($q=p+1$%
). Magnetic translations on the $i$-layer along the two cycles $A$ and $B$
are described by exponential of differential operators acting on the $w$
dependence of the characters. In the bilayer system the states belong to the
$1/2$ representation of the $su(2)$ pseudospin group. The TM on the torus
keeps track of these pseudospin configurations by the $w$ dependence of the
characters, whose charged and neutral components are described in terms of
the layers variables $w^{(1)}$, $w^{(2)}$ as $w_{c}=(w^{(1)}+w^{(2)})/2$ and
$w_{n}=(w^{(1)}-w^{(2)})/2$ respectively.

So the two configurations, given above, without tunneling are described on
the torus by the following translations on the charged and neutral $w$
coordinate. In the non crossed case (Fig. 2a) the potential $V_{AC}$ ($%
V_{BD} $) generates a translation along the first (second) layer, on the
variable $w^{(1)}$ $(w^{(2)})$, and it results $\Delta w_{c}\propto
V_{AC}+V_{BD}$ and $\Delta w_{n}\propto V_{AC}-V_{BD}$, while in the crossed
case (Fig. 2b) $\Delta w_{c}\propto V_{AD}+V_{BC}$ and $\Delta w_{n}\propto
V_{AD}-V_{BC}$. Let us point out that a purely neutral translation $w_{p}$
with $w^{(1)}=-w^{(2)}$ creates the topological defect (and relates the
edges switching to the large unbalance phenomenon predicted in \cite{edge1}%
). Finally, in the presence of localized tunneling ($t_{p}\neq 0$)\ between
the layers hybridization takes place. In fact that experimentally
corresponds to an equilibration process between the two edge states and
results into a breaking of the symmetry of the balanced system described by
the TM, due to the breaking of pseudospin symmetry. To take that into
account the boundary CFT technology was used in \cite{noi}, obtaining the
characters of the system in the presence of both tunneling and topological
defects.

Let us now discuss the transport properties in these unbalanced cases, by
describing the tunneling as a small perturbation to the TM, and focus our
attention to the terminal $AD$ in the crossed case. The working points are
different for the untwisted and the twisted configurations. In the first
case the term in eq. (\ref{tunnel}) for $t_{p}\ll 1$ is a weak perturbation
of the background characterized by $V_{AD}=V_{BC}=0$ while in the second one
it has $V_{AC}=V_{BD}=0$. The $I-V$ characteristics depends strongly on
that. We obtain a different conductance for the two cases. In particular for
TBC in the absence of an in plane magnetic field the driving voltage $V_{AD}$
puts the bilayer edges at different chemical potentials and then the ratio
of the $AD$ terminal current to $V_{AD}$ is equal to the Hall conductance $%
\sigma _{H}=\frac{e^{2}}{2h}$ of the single layer. For general filling $\nu =%
\frac{1}{p+1}$ the Hall conductance for single layer is expected to behave
as $\sigma_{H}=\frac{e^{2}}{2(p+1)h}$. Notice that the values of the slope
just obtained refer to the layers population condition $g=\nu -g=\frac{1}{%
2(p+1)}$ (balanced case).

Conversely, when the two layers are coupled via the in plane magnetic field,
the tunneling of the charge carriers results into a loss in the $AD$
terminal current. The net result is a negative contribution to the current
which adds to the previous term, producing a total $AD$ terminal current,
which for $p=0$, can be exactly evaluated in a similar way as in \cite
{fendley}, obtaining:
\begin{equation}
I_{AD}\left( V_{AD}\right) =\frac{e^{2}V_{AD}}{2h}-\frac{eT_{B}}{h}\arctan
\frac{eV_{AD}}{2T_{B}},
\end{equation}
where $T_{B}=C_{1}t_{P}^{1/\left( 1-\nu \right) }$ is the analogue of the
Kondo temperature, depending on the external parallel magnetic field, $C_{1}$
is a non-universal constant and $\nu $ is the filling (see Fig. 3). In this case $\nu =%
\frac{1}{2}$, for the single layer, but the argument can be generalized to a
wide class of fillings.
\begin{figure}[ht]
\centering\includegraphics*[width=0.7\linewidth]{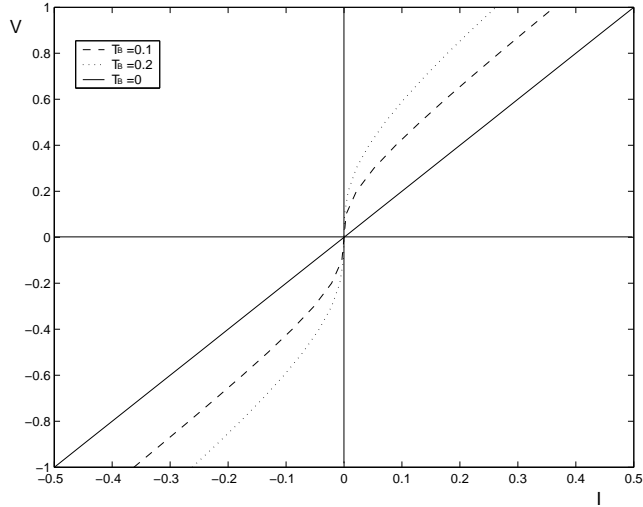}
\caption{I-V characteristics of the bilayer system in the twisted sector for
different values of $T_{B}$.}
\label{figura3}
\end{figure}

The non linear behavior of the tunneling characteristics follows
by standard analysis (ref. \cite{fendley}). Indeed for $T_{B}=0$
the characteristics has a linear behavior as for the transport in
a single layer (see Fig. 3). Moreover in plane magnetic field
removes the twist (topological defect) and re-establishes the
non-linear structure characterizing the tunneling phenomenon. Let
us notice also that our system is spinless (or fully polarized)
while the experimental results in \cite{deviatov1} are obtained
for spin resolved systems. Therefore we reproduce only the
negative branch of the curves given in \cite {deviatov1}. No gap
is obtained for positive $V_{AD}$.

In conclusion the presence of topological defects in a double layer induces
flux fractionalization described by the special $w_{p}$ translation and is
responsible for linear conduction between different edges with a quantized
value of the slope. We point out that the evidence of topological defects,
resulting from TBC, is theoretically indispensable for the consistency of
our CFT approach to the QHE. It is implied by the $m$-reduction technique.

At the moment an application of the above ideas to the spin-1/2 linear
chains with Mobius topology is under study.

\end{document}